# EFFECT OF WEATHER CONDITIONS ON FSO LINK

Saumya Borwankar
*Electronics and Communication Department*
*Institute of Technology, Nirma University*
Ahmedabad, India
17bec095@nirmauni.ac.in

Dhruv Shah
*Electronics and Communication Department*
*Institute of Technology, Nirma University*
Ahmedabad, India
17bec097@nirmauni.ac.in

*Abstract*— **Free Space Optics (FSO) is a developing technology for Line of Sight communication that uses light propagation in free space that provides various advantages like high bandwidth, high data rate, ease of installation, free licensing and secure communication. Thus, FSO is a developing technology that can be used in numerous applications for Line of Sight Communication. But the diverse effects like attenuation on FSO communication link due to environmental factors and weather conditions like fog, rain, dust, sand storms, clouds, temperature and the other factors like range, effects of physical obstructions are an essential topic for study which is discussed in this paper. We have done the simulation for the effects of fog and rain on the FSO communication link in Opti system software [1]. This is submitted in leu of FOC assignment at Nirma University.**

## I. INTRODUCTION

In this digital era, requirements for high data-rate in different applications are increasing day by day. So, there is a need for a system that can provide a solution with low cost implementation, reliable results and secure transmission of data. Free Space Optical (FSO) link is a perfect system that can provide speed, security and low-cost implementation for Line of Sight communication. FSO can provide such high data-rate in Gbps and can provide secure transmission as we know that optical communication provides a superior secure transmission then other types of transmissions. FSO can be implemented in various application since it is having major advantages of huge bandwidth, secure transmission, high data-rate, low cost implementation, free licencing and many more. FSO communication technique is used in inter-satellite communication, inter-building communication over a line of sight range, point to point or point to multipoint communication for short distance data transmission. But it is also important to analyse and discuss various diverse effects which can disrupt or make FSO communication less reliable so that as per the analysis we can find a solution and make this technology advance and full proof. The biggest challenge for the FSO communication is the effects of different weather conditions on transmission. We have to consider the different effects of different weather conditions since FSO communication takes place in free space. So, when there are adverse weather conditions, FSO becomes unreliable for data transmission. This is a major demerit over various merits of FSO communication. So, before setting up a link of FSO communication, for a long time period different effects of different weather conditions like fog, rain, sand storm, extreme temperature should be examined with the real implementation parameters like range, input power, optical source, wavelength and various other parameters. The biggest challenge with the FSO communication link is the effect of weather conditions, especially fog [2]. Because the fog particles have approximately the same size as the wavelength at which the FSO transmission takes place, which is 1550 nm. FSO link is prone to the light scattering, absorption and scintillation. So, we have done simulation in Opti system for practical implementation parameters for diverse weather conditions fog and rain for FSO link.

## II. DESCRIPTION OF FSO LINK

There is the description of the basic technology of FSO transmission link and the effects of diverse weather conditions - Rain and Fog on FSO transmission link.

### A. Technology of FSO Link

Working Principle of FSO is the same as optical fiber communication. But in FSO, we do not require optical fiber. Instead, we use free space as a channel. For a line of sight communication with FSO, we convert incoming data bits '1' and '0' into pulses of invisible light by optical source- LED or Laser. Then the transceiver transmits the aimed optical pulses with narrow beam width into air. There is another transceiver working as receiver inside the line of sight range which receives the optical pulses with the lenses and using avalanche photodiode or phototransistor the received light pulses are converted back into bits '1' and '0' and then are connected to a network. This optical transmission takes place at 1550 nm since at this wavelength the attenuation is lowest. The frequency spectrum is in GHz hence the available bandwidth is quite large so we can also have high data-rates for the transmission. Laser source is preferred over LED source since the former is having high reliability. The transmission can take place for a distance of a few kilometres. The range varies according to various parameters of free space at different times. There are several advantages like narrow beam width, no side lobes, secure transmission, huge bandwidth upto 2.5 GHz, high data-rate, low cost and less complex implementation and free licencing.[3]

### B. Effects of Weather Conditions on FSO Link

Different weather conditions like fog,rain,sandstorm,snow,low-clouds and different environmental parameters like varying temperature, air refraction index, density and also various pollution particles reduce the visibility and generate adverse effects especially attenuation of optical pulses at different intensity in FSO link and hence degrade the signal quality, making the communication unreliable. So long time analysis for the effects of these various parameters have to be done with real implementation parameters without actually setting up a FSO link. With the use of Opti system software, we simulated the FSO link with implementable actual parameters and analysed the results for adverse weather conditions - Fog and Rain. Apart from weather conditions, various physical and environmental parameters also attenuate the optical pulses.[4]

 1) *Effects of Fog on FSO Link:* Fog is the major factor for the degradation of signal quality since the size of fog particles is nearly the same as the wavelength of a carrier optical signal which is 550 nm. So, the attenuation of signal is very large than the attenuation due to other parameters. More attenuation decreases the range significantly so the receiver outside the degraded range cannot receive the optical pulses and hence lead us to the high BER rate or broken communication link. Now, if we consider the degraded range of 300 m and since the attenuation due to fog is quite large so we consider the attenuation factor of 100 dB/km, then the Q factor is found out to be of value 13. As we can see that the Q factor is quite low, so the energy loss is high and the pulse degrades rapidly. Eye diagram for the effects of fog on FSA link considering these parameters are attached and discussed in the Results section.[5]

 2) *Effects of Rain on FSO Link:* The impact on degradation of optical pulses due to rain is less than the effects of fog because the particles of rain drops are quite big in size than the wavelength. In case of snow-storms, the effect is less than the effects of rain because the size of snow particles are quite bigger than the wavelength. But due to rain, scattering phenomena of optical pulses happen and hence degrade the optical signal. We have to take the rain-rate into consideration because as rain-rate increases the scattering increases which causes degradation of optical signals. We have taken rain rate of 25 mm/hour and attenuation coefficient of 6 dB/km ( less than the coefficient of fog ) and we have gained Q factor of 58, which is quite bigger than the Q-factor due to fog which was 13. So, we can say that the impact of rain on the FSO link is quite less than the impact of fog on the FSO link. So, we can say that the ber value for same snr for rain will be less compared to the ber value for fog and the range degraded will be less in case of impact of rain than the impact of fog. Eye diagram for the impact of rain on FSO

link for the parameters and their values mentioned above is attached and discussed in Results section.[6]

## III. SIMULATION SETUP

In Fig.1 a designed system of FSO in Opti system is shown. Here, block 1 shows a laser source having wavelength of 1550nm because atmospheric attenuation produces less effect at this wavelength. Block 2 consists of a subsystem that has PRBS (Pseudo Random Bit Sequence) generator, NRZ pulse generator, low pass Bessel filter at the cut-off frequency of external bandwidth and Machzender modulator. In Fig.2 subsystem is shown. After that Fiber Space Optics channel is present which comprises a link range as per requirement, attenuation factor, Tx and Rx aperture diameter and beam divergence. Then the receiving part is given. Fourth block is an APD (Avalanche Photo Detector) having ionization ratio 0.9 and 10nA dark current which converts the optical signals into electrical signals. Fifth block is a low pass Bessel filter with cut-off frequency= 0.75*Bit rate and order of 4. Here the bit rate is taken as 10 Gbps.[7] At the end, the BER analyser is used to compute the eye diagram and Q-factor of the designated system.

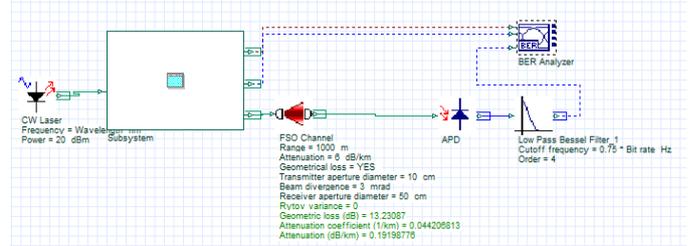
Fig. 1 Simulation model of FSO link

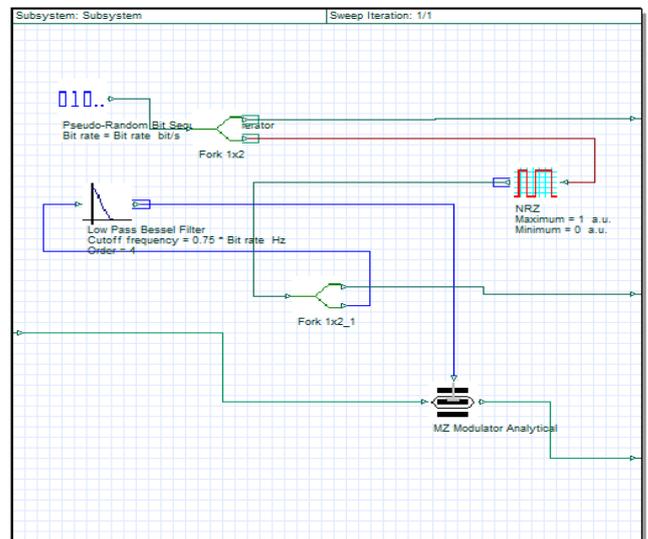
Fig. 2 Subsystem design

### A. Important Simulation Parameters

The parameters of all blocks for the rain and fog are as shown in Table I.

| Parameters | Value |
|------------|-------|
| CW laser | Frequency: 1550nm |
|  | Bandwidth: 5 MHz |
| FSO | For Rain |
|  | Attenuation:6 dB/km |
|  | link range: 1km |
|  | For Rain |
|  | Attenuation:100 dB/km |
|  | link range: 300m |
| APD photodiode | Responsivity: 1 A/W |
|  | Dark current: 10nA |
|  | ionization ratio: 0.9 |
| Low Pass Bessel Filter | Cut off frequency: 0.75*Bit rate |
|  | Order:4 |

| Parameters | Value |
|------------|-------|
| Data Rate | 10Gbps |
| Input power | 20dBm |
| Sequence length | 128 bits |
| Samples per bits | 64 |
| Number of samples | 8192 |
| Beam divergence | 3mrad |

### B. Important Results Parameters

Here, three important results parameters are discussed which are obtained at receiver. They are discussed below.

1) *Link Margin*: Link margin is calculated by observing received signal power at the receiver side and it can affect the quality and performance of a FSO link. The equation for the link margin (LM) is as below.

$$LM = 10 \log P_R / S \qquad (1)$$

In above equation, $P_R$ is received signal power and S is the receiver sensitivity which is a constant manufacturing value ranging between -20 to -40 dBm. As per the simulation parameters and reference parameters we obtained value of $P_R$ for rain and fog which are -5.066 dBm and -19.099 dBm respectively. And we assume the receiver sensitivity of -20 dBm. So, by using the equation given above the values of link margin (LM) are,

a) For rain, LM = -5.9636
b) For fog, LM = -0.2001

2) *Geometric Attenuation*: When an optical pulse is passing through the air, geometric attenuation causes divergence of the optical beam. It also implies the performance analysis of FSO link. Geometric attenuation is given by the equation below and measured in dB

$$A_{geo} = [d_{rx} / d_{tx} + \Theta l]^2 \qquad (2)$$

In this equation, $d_{rx}$ and $d_{tx}$ are receiver and transmitter aperture diameters in cm and $\Theta$ is divergence angle and '$l$' is link length. Divergence angle is 3 mrad and and link length is 1 km and 300 m for rain and fog respectively as per the Table II. The geometric attenuation for rain and fog are displayed in Fig. 4 and Fig. 6 respectively with the values needed in equations.

3) *Atmospheric Attenuation*: Attenuation occurs due to aerosols present in the channel is termed as atmospheric attenuation. Due to atmospheric attenuation, signal gets distorted and results in scattering, absorption and diffraction. So, in these diverse conditions, under the atmospheric attenuation, as our signal propagates through the channel, power and intensity decrease. Total transmittance of the atmosphere at the optical wavelength is given by the Beer-Lambert law equations [8] as below.

$$\tau(\lambda, d) = P(\lambda, d) / P(\lambda, 0) = e^{-\gamma(\lambda)d} \qquad (3)$$

In this equation, $P(\lambda, d)$ denotes the power at distance d for a particular wavelength, $P(\lambda, 0)$ denotes the power at transmitter without the effects of attenuation and $\gamma(\lambda)$ denotes attenuation coefficient per length unit (dB/km) which is found while doing simulation and is shown in Fig. 4 and Fig. 6 for rain and fog respectively. So, from this equation the values for the transmittance at distance of 1 km for rain and 300 m for fog are,

a) For rain, $\tau(\lambda, d) = 0.8253$
b) For fog, $\tau(\lambda, d) = 0.9440$

We can also find out the intensity which is decreased at distance d due to the attenuation which was caused by different molecules and aerosols of the channel. Intensity at distance d from the source can be found out by the Beer's law which is shown in the equation below.

$$I = I_0 e^{(-\lambda d)} \qquad (4)$$

Where, I and $I_0$ are detected intensity at distance 'd' and intensity at transmitter respectively and '$\lambda$' is attenuation coefficient which is found and displayed in Fig. 4 and Fig. 6.

## IV. RESULTS

Eye pattern or eye diagram is basically an oscilloscope display. In eye pattern, a digital signal from the receiver is continuously sampled and fed to the vertical input and the horizontal sweep is triggered by the data-rate. With the help of eye diagrams, we are able to evaluate the effects of channel noise and Intersymbol Interference (ISI) on the performance of a baseband pulse-transmission system. The eye patterns for various weather conditions for defined parameters value as discussed in the topics-Effects of fog on FSO link and Effects of rain on FSO link are simulated and presented below.

## 1. Effects of Rain on FSO Link

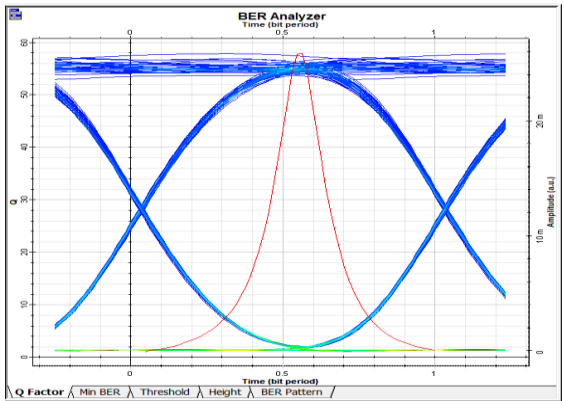

Fig. 3 Eye diagram for simulation setup of Rain Attenuation

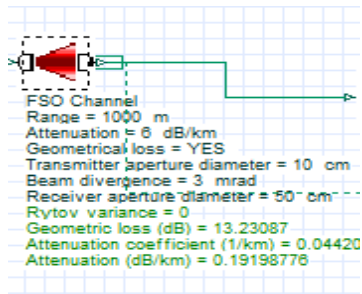

Fig. 4 FSO channel parameters for rain attenuation

## 2. Effects of Rain on FSO Link

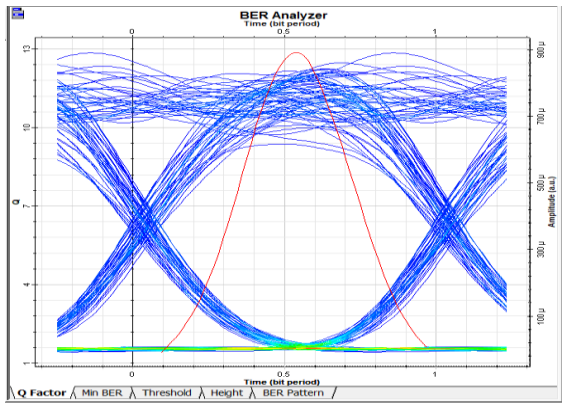

Fig. 5 Eye diagram for simulation setup of fog attenuation

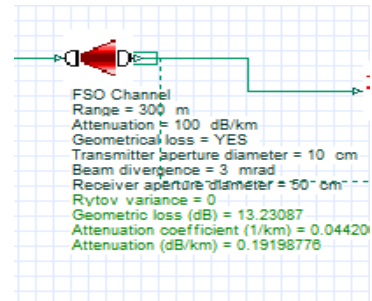

Fig. 6   FSO channel parameters for fog attenuation

As we can see in Fig. 3 and Fig. 5 for rain and fog respectively that the eye diagram for rain in Fig. 3 is quite better than the eye diagram for fog in Fig. 5.

## V. FUTURE SCOPE

Free Space Optics (FSO) is a line of sight communication system. A further detailed study can be made on how to remove the adverse effects of different weather conditions and environmental factors. And by doing so a vast field of applications can be discovered where we can apply this FSO system for a fast, secure and reliable communication technique.

## VI. CONCLUSION

Free Space Optics (FSO) is a line of sight communication system which can provide numerous advantages of high bandwidth, high data-rate, secure communication and many more. But the adverse effects of weather and environmental factors make this communication technique less reliable. So, we have analysed adverse effects of fog and rain on a FSO system and have acquired various parameters like link margin and geometric attenuation. We have also acquired eye diagrams for these effects of fog and rain on a FSO system in Opti system software successfully.